\documentclass[a4paper,11pt]{article}
\usepackage{jcappub}
\usepackage{url}

\title{Detecting the neutrino mass and mass hierarchy from global data}

\author[a]{Wenxue Zhang}
\author[a]{En-Kun li}
\author[a]{Minghui Du}
\author[a]{Yuhao Mu}
\author[a]{Shouli Ning}
\author[a]{Baorong Chang}
\author[a,1]{Lixin Xu\note{Corresponding author.}}
\emailAdd{lxxu@dlut.edu.cn}

\affiliation[a]{Institute of Theoretical Physics, School of Physics, Dalian University of Technology, \\ Dalian, 116024, China}

\abstract{
   In this paper, we have constrained the neutrino mass and mass hierarchy in the $\Lambda$CDM cosmology with the neutrino mass hierarchy parameter $\Delta$, which represents different mass orderings, by using the {\it Planck} 2015 + BAO + SN + $H_{0}$ data set, together with the neutrino oscillation and neutrinoless double beta decay data. We find that the mass of the lightest neutrinos and the total neutrino mass are no more than $0.035$\text{eV} and $0.133$\text{eV} at $95\%$ confidence level, respectively. Comparing the result of our joint analysis with that obtained using cosmological data alone, we find that, by adding the neutrino oscillation and neutrinoless double beta decay data, the tendency for normal hierarchy has increased a lot. By means of importance sampling, three other priors are taken into account, i.e., the flat logarithmic prior on the absolute value of the neutrino hierarchy parameter $\Delta$, the flat linear prior on the total neutrino mass $\Sigma m_{\nu}$, and the flat logarithmic prior on $\Sigma m_{\nu}$. We find that the preference for the normal hierarchy is in agreement whatever what kinds of priors we choose. Finally, we make a Bayesian model analysis about four priors and we find that flat-linear and the flat logarithmic priors on $\Sigma m_{\nu}$ are the most favored priors.

}
\begin{document}
\maketitle
\flushbottom

\section{Introduction}
\label{sec:intro}

Modern neutrino oscillation experiments have confirmed that neutrinos have masses, and the solar and atmospheric neutrino experiments suggest that there are at least two of neutrinos mass eigenstates are non-zero.
This implies the first departure from the Standard Model of Particle Physics.
Science the neutrino oscillation experiments are only sensitive to the squared mass differences, scientists can only confirm that $\Delta m^2_{21} = m_2^2 - m_1^2>0$ and $|\Delta m^2_{31}| = |m_3^2 -m_1^2|>0$, where $m_i$ is the three mass eigenstates \cite{fukuda1998evidence, Mcdonald:2016wje, Kajita:2016gnc, de1708status}.
Until now, the sign of $\Delta m^2_{31}$ is still unknown, this lead to two possible orderings of neutrino masses, i.e. the normal hierarchy (NH), which means $m^2_{31}> 0$ and $m_1<m_2<m_3$; and the inverted hierarchy (IH), which means $m^2_{31}<0$ and $m_3<m_1<m_2$ \cite{Vagnozzi:2017ovm}.
Though the future terrestrial experiments, which are designed to exploit the matter effects in Earth, will be dedicated to figure out the sign of $\Delta m^2_{31}$ by using long baseline accelerators \cite{Acciarri:2015uup} and atmospheric experiments \cite{Aartsen:2014oha, Adrian-Martinez:2016fdl}, the cosmological observations can shed light on determining the absolute values of neutrino masses due to the gravity is much sensitive to the mass distribution.

The relic neutrino have significance effects on the cosmological evolution, both at background and perturbation level, so that cosmological observations can be used to constrain the neutrino properties, in particular their masses (see \cite{dolgov2002cosmological, Lesgourgues:2018ncw, Kamau2004Neutrino, lesgourgues2006massive, Wong2011Neutrino, lesgourgues2012neutrino, Abazajian:2016hbv, lesgourgues2014neutrino} for detail). The absolute mass of neutrinos can be determined by observing the cosmic microwave background radiation and large-scale structure.
Using the cosmic microwave background (CMB) data alone, authors in \cite{Lattanzi:2017ubx} find that $\sum m_\nu < 0.7$ \text{eV} ($95\%$C.L.). 
This limit has been brought down to $\sum m_\nu < 0.17$\text{eV} ($95\% $C.L.) by adopting the combination of \textit{Planck2015} TT,TE,EE+lowP and baryon acoustic oscillation (BAO) data \cite{Ade:2015xua}.
The latest data published by $\textit{Planck2018}$ \cite{Aghanim:2018eyx} shows that the maximum sum of neutrino masses is $\Sigma m_{\nu}<0.24$eV ($95\%$confidence, TT,TE,EE+lowE+lensing). 
And the measurements of the CMB anisotropies together with the BAO (\textit{Planck2018} TT,TE,EE+lowE+lensing+BAO) find that the neutrino mass is tightly constrained to $\sum m_\nu < 0.12$ eV \cite{Aghanim:2018eyx}, which is very close to the lower limit of the total neutrino mass in IH case. 
Next-generation CMB and large scale surveys will significantly tighten the present cosmological limit on $\sum m_\nu$ \cite{Archidiacono:2016lnv, Hamann:2012fe} and will likely have the capability to figure out the mass ordering.

In Refs. \cite{Caldwell:2017mqu, Gariazzo:2018pei}, the authors present a full Bayesian analysis utilizing the combination of current neutrino oscillation, neutrinoless double beta decay and Cosmic Microwave Background observations, and find that they only moderately favor the normal ordering. Authors in Refs. \cite{Xu:2016ddc} propose a neutrino mass hierarchy parameter and find that the normal hierarchy is slightly favoured by using the data combination Planck 2105 LowTEB, TT, TE, EE + BAO DR12 + JLA SN + HST 2016.

Another way used to determine the absolute mass of the neutrino comes from accurately measuring the tritium $\beta$ decay spectrum. 
Tritium decay of $\beta$ $(^{3}_{1}H\rightarrow^{3}_{2} He + e^{} + \bar{\nu_e})$ in the process of final state to obtain the biggest kinetic energy electron neutrino absolute quality. By accurately measuring the energy spectrum of the electron, we can observe the kinematic effects of the mass of the neutrino. The upper limit of the effective mass of neutrinos given in the current experiment is $2.2\text{eV}$ ($95\%$C.L.)\cite{Lobashev1999Direct,Weinheimer2000High}, and the accuracy of the forthcoming $\rm{KATRIN}$ experiment can reach $0.2 \text{eV}$\cite{Angrik:2005ep}.
Neutrinoless double beta decay is a hypothetical nuclear process beyond the Standard Model in which two neutrons undergo $\beta$ decay simultaneously without the emission of neutrinos and the total lepton number is violated by two units \cite{GomezCadenas:2011it}. Observation of this decay would probe whether neutrino is Majorana particle and provide us with precious information on the neutrino mass scale and ordering \cite{DellOro:2016tmg}. Next-generation experiments of neutrinoless beta decay could have the potential to solve the hierarchy pattern \cite{KamLAND-Zen:2016pfg, Gerbino:2015ixa}.

Following the method raised from \cite{Caldwell:2017mqu, Gariazzo:2018pei}, in oue paper, we will use global observational data, which include data from several cosmic observational measurements, the neutrino oscillation experiments and neutrinoless double beta decay, to constraint neutrino mass and mass hierarchy.
The rest of the paper is organized as follows. In section \ref{sec:methda}, we will describe the methodology and parametrizations we use in this paper.  In section \ref{sec:result}, we will report our results and make some discussions in depth about the fitting results. Finally, a brief summary will be presented in section \ref{sec:conclu}. In Appendix \ref{app:data}, we will present the details of experimental constraints, which include neutrino oscillation, neutrinoless double beta decay and cosmological data.

\section{Method and data}
\label{sec:methda}

The baseline model that we will extend to study various neutrinos properties is the $\Lambda$CDM model, described by the six usual parameters, and we add one hierarchy parameter:
    \begin{equation}
        \{ \omega_b, \omega_c, 100 \theta_{MC}, \tau, n_s, \ln[10^{10} A_s], \Delta \},
        \label{eq:parameters}
    \end{equation}
where $\omega_b = \Omega_b h^2$ and $\omega_c = \Omega_c h^2$ are the present-day baryon and cold dark matter energy densities respectively, $\theta_{MC}$ is the ratio between the sound horizon and the angular diameter distance at the time of last-scattering, $\tau$ is the Thomson scattering optical depth due to the reionization, $n_s$ is the spectral index of scalar power spectrum and $A_s$ is the amplitude of the power spectrum of primordial curvature perturbations, and $\Delta=\frac{m_3-m_1}{m_1+m_3}$ is the dimensionless neutrino mass hierarchy parameter proposed by Ref.\cite{Xu:2016ddc}

The introduction of hierarchy parameter can save a lot of computing resources and the sign of the hierarchy parameter can directly indicate the condition of the ordering, where the positive sign denotes the normal mass hierarchy and negative sign denotes the inverted mass hierarchy \cite{Li:2017iur}. Using the hierarchy parameter $\Delta$, the three neutrino mass eigenvalues can be rewritten as following:
 \begin{align}
        &m_1 = \frac{1-\Delta}{2} \sqrt{\left| \frac{\Delta m_{31}^2}{\Delta} \right|},
        \label{eq:m1} \\
        &m_2 = \sqrt{\frac{(1-\Delta)^2}{4} \left|\frac{\Delta m_{31}^2}{\Delta}\right| +\Delta m_{21}^2},
        \label{eq:m2} \\
        &m_3 = \frac{1+\Delta}{2} \sqrt{\left|{ \frac{\Delta m_{31}^2}{\Delta} }\right|}.
        \label{eq:m3}
    \end{align}
And the total neutrino mass will be given by
 \begin{equation}
        \sum m_\nu=\sqrt{\left|\frac{\Delta m_{31}^2}{\Delta}\right| } + \sqrt{\frac{(1-\Delta)^2}{4} \left|\frac{\Delta m_{31}^2}{\Delta} \right|+\Delta m_{21}^2} .
        \label{eq:summ}
    \end{equation}

\begin{table}[t]
\centering
\begin{tabular}{|c|c||c|c||c|c|}
\hline
\multicolumn{2}{|c||}{Cosmological} & \multicolumn{2}{|c||}{$0\nu\beta\beta$} & \multicolumn{2}{|c|}{Neutrino mixing} \\
\hline
Parameter & Prior & Parameter & Prior & Parameter & Prior \\
\hline
$\Omega_b h^2$ & 0.019 -- 0.025 & $\alpha_2$ & 0 -- $2\pi$  & $\sin^2 \theta_{12}$  & 0.1 -- 0.6 \\
$\Omega_c h^2$ & 0.095 -- 0.145 & $\alpha_3$ & 0 -- $2\pi$  & $\sin^2 \theta_{13}$  & 0.00 -- 0.06 \\
$\Theta_s$ & 1.03 -- 1.05  & $\mathcal{M}^{0\nu}_{^{76}{\rm Ge}}$ & 4.07 -- 4.87   & $\sin^2 \theta_{23}$ & 0.25 -- 0.75 \\
$\tau$     & 0.01 -- 0.4   & $\mathcal{M}^{0\nu}_{^{136}{\rm Xe}}$ & 2.74 -- 3.45  & $\Delta m_{21}^2[10^{-3} \text{eV}^2]$ & 0.001 -- 1 \\
$n_s$      & 0.885 -- 1.04 &   &                                              & $\Delta m_{31}^2[10^{-3} \text{eV}^2](\text{NH})$ & 0.2 -- 7 \\
$\ln[10^{10} A_s]$    & 2.5 -- 3.7  &  &                                      & $\Delta m_{32}^2[10^{-3} \text{eV}^2](\text{IH})$ & -7 -- -0.2 \\
$\Delta$ &-1 -- 1 & &   &  & \\
\hline
\end{tabular}
\caption{Cosmological parameters, $0\nu\beta\beta$ parameters and neutrino mixing angles used in the analysis, with the adopted priors.}
\label{tab:commonParams}
\end{table}

The standard neutrino oscillations indicate that the neutrino flavour eigenstates $\nu_\alpha$ ($\alpha= e,\,\mu,\,\tau$) are quantum superpositions of three mass eigenstates $m_i$ through a unitary matrix  $U_{\alpha i}$, which are called the Pontecorvo-Maki-Nakagawa-Sakata (PMNS) matrix :
\begin{equation}
  |\nu_\alpha \rangle = \sum\limits_{i} U^{*}_{\alpha i} |\nu_i \rangle,
\label{eq:nu}
\end{equation}
where $U_{\alpha i}$ can be parameterized by three mixing angles($\theta_{12},\,\theta_{23},\,\theta_{13}$), one Dirac CP violating phase ($\delta$) and two Majorana phases ($\alpha_{21},\,\alpha_{31}$) \cite{Lesgourgues:2018ncw}. Oscillation phenomena are insensitive to the two Majorana phases, meanwhile, the CP violating phase have not been strongly constrained in the neutrino oscillation measurements and it does not affect cosmological or $0\nu\beta\beta$ observation, so in the neutrino oscillations we only consider three mixing angles and use them as physical parameters in the form $\sin^2 \theta_{12},\,\sin^2 \theta_{13},\,$ and $\sin^2 \theta_{23}$ \cite{Fogli:2005cq}. The probablity of neutrino oscillations always depends on $\Delta m^{2}$, i.e. on the squared mass difference, so we also take $\Delta m^{2}_{21}$, $\Delta m^{2}_{31}$ (NH) and $\Delta m^{2}_{32}$(IH) into account.

The actual parameter measured in any neutrinoless double beta decay experiment is the half-life $T^{0\nu}_{1/2}$, which is associated with  the effective Majorana mass, 

\begin{equation}
\frac{1}{T^{0\nu}_{1/2}}=G_{0\nu}\,{\left| \mathcal{M}^{0\nu} \right|}^{2}\,\left( \frac{\left|m_{\beta\beta}\right|}{m_e} \right)^{2}\,,
\label{eq:thalf}
%\label{eq:thalf}
%\label{eq:meff}
\end{equation}
and the effective Majorana mass is related to the neutrino mass eigenvalues
\begin{equation}
m_{\beta\beta} = \left|\sum_{k} e^{i\alpha_k}\,U_{\rm PMNS,ek}^2\, m_k\right|\,,\label{eq:meff} 
\end{equation}
%\begin{equation}
%m_{\beta\beta} = \frac{m_e}{\mathcal{M}^{0\nu}}\sqrt{G_{0\nu}T^{0\nu}_{1/2}}~,
%\label{eq:thalf}
%\end{equation}
where  $G_{0\nu}$ is a phase-space factor dependent on the charge, mass and available energy of the process, and we adopt the value $0.623\times 10^{-14}$ yr$^{-1}$ for ${}^{76}{G}_{0\nu}$ and $4.31\times 10^{-14}$ yr$^{-1}$ for ${}^{136}{G}_{0\nu}$ \cite{Rodejohann:2011mu}, $m_e$ is the electron mass. $\mathcal{M}^{0\nu}$ is the nuclear matrix element(NME), following the reference \cite{giuliani2012neutrinoless}, we adopt the range [4.07,4.87] for ${}^{76}{Ge}$ and [2.74,3.45] for ${}^{136}{Xe}$. $\alpha_{k}$ ($k=1,2,3$) are the Majorana phases. One of the phases can always be rotated away so we can assume that $\alpha_1 = 0$ by convention. The neutrino Majorana phases play no role in neutrino oscillation process, however, they are crucial for $0\nu\beta\beta$ experiments. Following works \cite{Gerbino:2015ixa, Gerbino:2016ehw}, we adopt flat prior in the range $[0,\,2\pi]$ for the Majorana phases $\alpha_2$ and $\alpha_3$.

We constrain the neutrino mass and mass hierarchy in the $\Lambda$CDM cosmology model with the neutrino mass hierarchy parameter $\Delta$ by using all the {\it Planck} 2015 + BAO + SN + $H_{0}$ data, together with the global fit of neutrino oscillation and neutrinoless double beta decay, and we have taken the the likelihood of aforementioned data  as the part of the global likehood function $L\propto e^{-\chi^2/2}$, then we can get $\chi^2$ 
\begin{eqnarray}
\chi^2_{\rm global}=\chi^2_{\rm cosmos}+\chi^2_{\rm osc}+\chi^2_{0\nu\beta\beta},
\end{eqnarray}
where the seperate likehoods of the current cosmological datasets, neutrino oscillation and $0\nu\beta\beta$ used in this paper are shown in the Appendix \ref{app:data}. All adopted parameters prior ranges are listed in Table.~\ref{tab:commonParams}. Our constraints are based on the Monte Carlo Markov Chain package \textbf{CosmoMC} \cite{Lewis:2002ah, Lewis:1999bs}, and we have generated 8 MCMC chains.

To measure a model's performance, the Bayesian evidence is always mentioned. For a continuous parameter space $\Omega_{\mathcal{M}}$ and aimed model $\mathcal{M}$,  the Bayesian evidence is given by:
\begin{equation}
p(d \mid \mathcal{M}) = \int_{\Omega_\mathcal{M}} p(d\mid \theta, \mathcal{M}) p(\theta\mid\mathcal{M}) d\theta ,
\label{bayesian}
\end{equation}
where $d$ is the datasets, $\theta$ is a set of parameters, $p(d\mid \theta, \mathcal{M})$ is the likelihood, and $p(\theta\mid\mathcal{M})$ is  the prior.
By using Bayes' theorem, the model posterior probability is given by
\begin{equation}
p(\mathcal{M} \mid d ) \propto  p(\mathcal{M}) p(d \mid \mathcal{M}) ,
\label{baye}
\end{equation}
where $p(\mathcal{M})$ is the prior probability and related to the model itself. When we compare two models, they have identical prior probability $p(\mathcal{M}_{1})=p(\mathcal{M}_{2})$, and the ratio of the posterior probability can be written as 

\begin{equation}
\frac{p(\mathcal{M}_{1} \mid d )}{p(\mathcal{M}_{2} \mid d )}=B_{12} \frac{p(\mathcal{M}_{1})}{p(\mathcal{M}_{2})},
\label{bayeratio}
\end{equation}
where $B_{12}$ is the Bayes factor and it represents the ratio of the evidence of two models:

\begin{equation}
B_{12}=\frac{p(d \mid \mathcal{M}_{1} )}{p(d \mid \mathcal{M}_{2} )}.
\label{bayefactor}
\end{equation}

Bayes factors are usually interpreted against the Jeffrey's scale for the strength of evidence \cite{Trotta:2008qt}, $\left|\ln{B_{12}}\right|<1$ is regarded as inclusive, $\left| \ln{B_{12}} \right| \in [1.0,2.5]$ is weak evidence, $\left| \ln{B_{12}} \right| \in[2.5,5.0]$ is moderate evidence, and $\left|  \ln{B_{12}} \right| >5.0$ is strong evidence.

The computation of the Bayesian evidence can take advantage of the MCMC chains to extract the parameters space of all kinds of datasets, and we use the code \textbf{MCEvidence}\footnote{\url{https://github.com/yabebalFantaye/MCEvidence}} to calculate the Bayesian evidence for different models \cite{Yang:2018xah}.

\section{Results and Analysis}
\label{sec:result}

\begin{table}[t]
\centering
\begin{tabular}{cc| cc}
\hline\hline Parameters & & mean with errors\\ 
\hline
$\sin^2\theta_{12}$ && $0.308^{+0.012+0.025+0.033}_{-0.012-0.023-0.030}$ \\
$\sin^2\theta_{13}$ && $0.0220^{+0.0008+0.0016+0.0021}_{-0.0008-0.0015-0.0019}$ \\
$\sin^2\theta_{23}$ && $0.520^{+0.050+0.081+0.094}_{-0.050-0.089-0.10}$ \\
$\Delta$ && $\Delta > 0.211 (95\% $C.L.)   \\
$\Delta m_{21}^2[10^{-5} \text{eV}^2]$ && $7.40^{+0.21+0.41+0.55}_{-0.21-0.41-0.54}$ \\
$\Delta m_{31}^2[10^{-3} \text{eV}^2]\text{(NH)}$ && $2.547^{+0.005+0.050}_{-0.069-0.11}\Delta m_{31}^2 > 2.40$ \\
\hline 
$m_{\nu,min} \text{eV}$ && $m_{\nu,min} < 0.035 (95\% $C.L.) \\
$\Sigma_{\nu}m_{\nu} \text{eV}$ && $0.082^{+0.004+0.051+0.093}_{-0.024-0.025-0.025}$ \\ 

\hline\hline 
\end{tabular}
\caption{Fitting results of based and derived parameters from the constraint of the data combination cosmological+neutrino oscillation+$0\nu\beta\beta$. }
\label{tab:results}
\end{table}

\begin{figure}[!htbp]
\centering
\includegraphics[width=12.0cm]{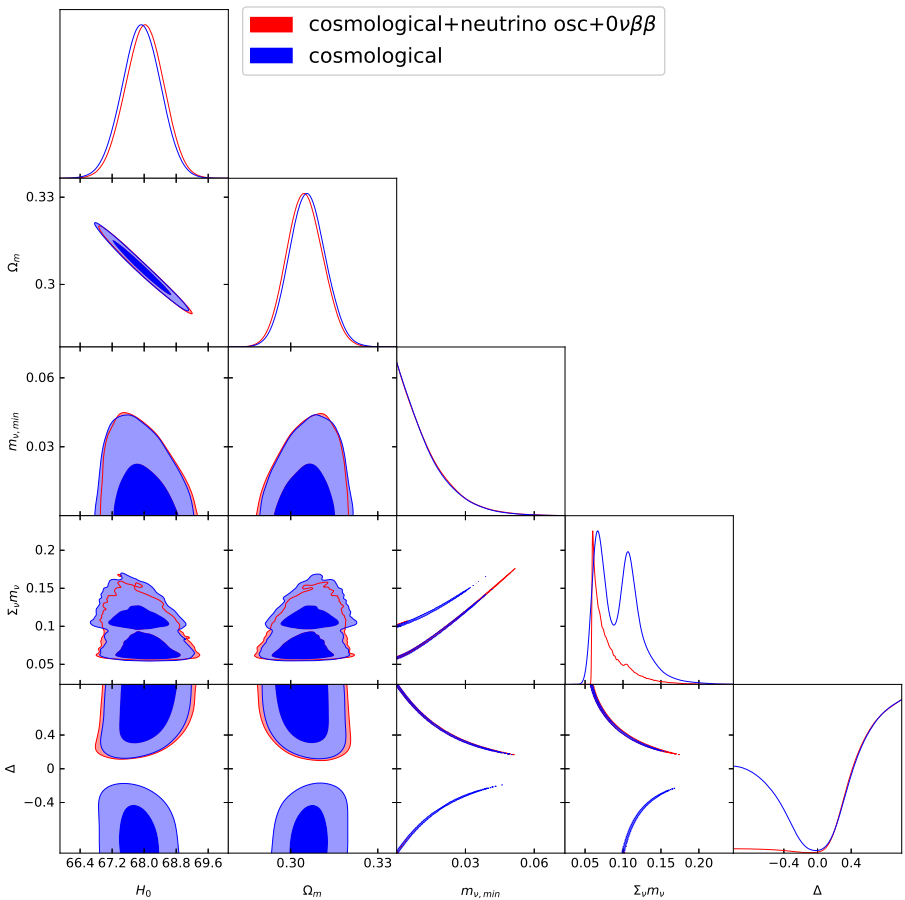}
\caption{The one-dimensional and two-dimensional joint, marginalized constraints from cosmological data (blue) and cosmological data + neutrino oscillation + $0\nu\beta\beta$ (red) for the relevant mass parameters with $68\%$ and $95\%$ confidence level. }
\label{fig:compare}
\end{figure}

Our constraint results are summarized in Table.~\ref{tab:commonParams} and Fig.~\ref{fig:compare}. The fitting results of basic and derived parameters from the constraint of the data combination cosmological+neutrino oscillation+$0\nu\beta\beta$ are shown in Table.~\ref{tab:commonParams}. Specially, from Table.~\ref{tab:commonParams}, one can find that $\Sigma m_{\nu} =0.082^{+0.004}_{-0.024}\text{eV} $, which is less than the minimal value of the total neutrino masses in IH case, suggesting IH case is ruled out.

Fig.~\ref{fig:compare}  displays the joint, marginalized constraints from cosmological data (blue) and cosmological data \cite{Xu:2016ddc} + neutrino oscillation + $0\nu\beta\beta$ (red). The $68\%$C.L. and $95\%$ C.L. contours for the interested parameters are shown. We find that, in both situations, $\Delta = 0 $ are strongly disfavored. When only using cosmological data, there are two peaks in the posterior distribution of hierarchy parameter, and the posterior at $\Delta>0$ is a little higher than $\Delta<0$, which means  NH is just slightly or weakly favoured  than IH. While using cosmological data + neutrino oscillation+$0\nu\beta\beta$, most of the posterior distribution of hierarchy parameter falls into the $\Delta > 0$ ranges, and a little in the $\Delta < 0$ intervals, which indicates that NH is evidently favoured in our results. Under our method, the combination of cosmological observational data with neutrino oscillation and neutrinoless double beta decay  data is intensely favour normal hierarchy.

\begin{figure}[!htbp]
\centering
\includegraphics[width=10.2cm]{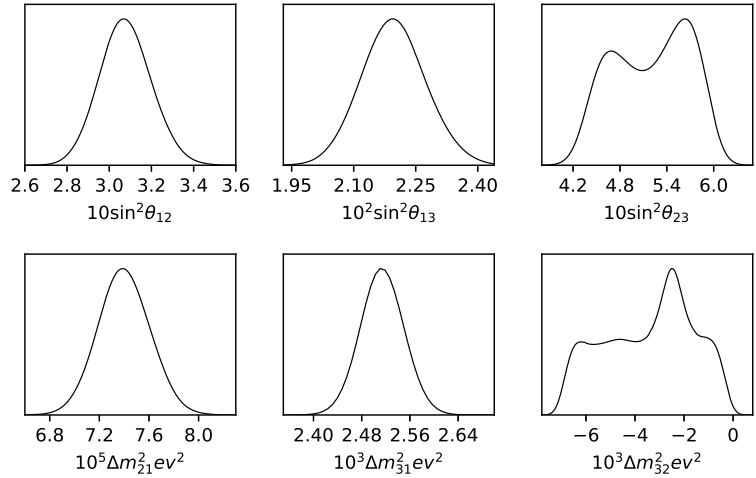}
\caption{The one-dimensional marginalized distribution for the relevant mixing angles(up) and for the squared mass differences (down).}
\label{fig:sinmasss}
\end{figure}

Since $\Delta$ covers the posibilities of both NH and IH, one can clearly find that there are two peaks of $\Sigma_{\nu}m_{\nu}$ when using cosmological data, but only one peak when using global data. The sum of total neutrino masses $\Sigma_{\nu}m_{\nu}$ at $95\%$ confidence level is no more than $0.147 \text{eV}$ and $0.133 \text{eV}$ for cosmological data and global data , respectively. And the upper limit of the lightest neutrino mass $m_{\nu,min}$ are $0.0341\text{eV}$, $0.0346\text{eV}$ for cosmological data and global data at $95\%$ confidence level. The one-dimensional posterior distributions of mixing angles and squared mass differences are shown in Fig.~\ref{fig:sinmasss}. 
Comparing to the results of reference \cite{Gariazzo:2018pei}, which provides Bayesian anlasis results on the neutrino mixing parameters from the global dataset including the \textit{Planck2015} CMB data, neutrino oscillation and neutrinoless double beta decay data, we see that our constraint on the mixing angles and squareed mass differences are consistent with their NH case results. However, the different is that our method gives a more smaller value of $\Sigma m_{\nu}$, which suggest NH case is prefered.
Recent global fits to data from the latest neutrino oscillation experiments,  which obtained constraints two different mass squared splittings: $\Delta m^2_{21}\approx 7.39^{+0.21}_{-0.20}\times 10^{-5}\ \text{eV}^2, |\Delta m^2_{31}|\approx 2.525^{+0.033}_{-0.031}\times 10^{-3}\ \text{eV}^2(1\sigma$ C.L.)\cite{Esteban:2018azc}. Our results, which $\Delta m^2_{21}\approx 7.40^{+0.21}_{-0.21}\times 10^{-5}\ \text{eV}^2, \Delta m^2_{31}$(NH)$\approx 2.547^{+0.005}_{-0.069}\times 10^{-3}\ \text{eV}^2(1\sigma$ C.L.), are in accordance with them constraint results in the two neutrino mixing parameter.

Due to the different priors may have a great influence on constraining the upper limit of the total neutrino mass \cite{Gariazzo:2018pei}. 
Now, we will compare the influences of different priors on $\Delta$ by means of importance sampling technique. Consider three kinds of priors: the flat logarithmic prior on the absolute value of the neutrino hierarchy parameter $\Delta$, the flat linear prior on the total neutrino mass $\Sigma m_{\nu}$, and the flat logarithmic prior on $\Sigma m_{\nu}$. The complete list of priors is reported in Table.~\ref{tab:priors} \cite{Li:2017iur}.
\begin{table}[t]
\centering
\begin{tabular}{cc|cc|cc|cc|}
\hline Parameters & & Prior && Range && Corresponding Prior on $\Delta$ \\ 
\hline
$|\Delta|$ && logarithmic && $[10^{-4}, 1]$ && $\propto 1/|\Delta|$ \\
$\Sigma m_{\nu}$ \text{eV} && linear && $[\Sigma, 7.50]$ && $\propto |d\Sigma m_{\nu}/d\Delta|$ \\
$\Sigma m_{\nu}$ \text{eV} && logarithmic &&  $[\Sigma, 7.50]$ &&  $\propto \frac{|d\Sigma m_{\nu}/d\Delta|}{\Sigma m_{\nu}}$  \\
\hline\hline 
\end{tabular}
\caption{Three kinds of priors, where $\Sigma =0.06$\text{eV} and $0.01$ \text{eV} for NH and IH, respectively. }
\label{tab:priors}
\end{table}

\begin{figure}[!htbp]
\centering
\includegraphics[width=7.2cm]{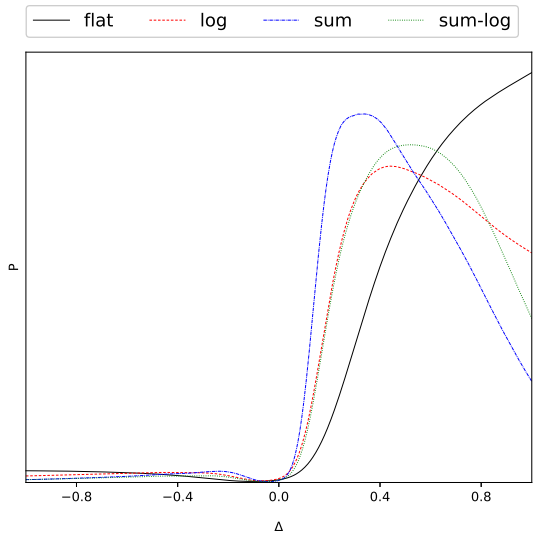}
\caption{Different importance sampling results for different priors. The balck-solid line is for flat-linear $\Delta$ prior, red-dashed line is for the flat-log $\Delta$ prior, blue-dashed line is for the flat-linear $\Sigma m$ prior, and green-dashed line is for the flat-log $\Sigma m$ prior.}
\label{fig:delta_prior}
\end{figure}

Fig.~\ref{fig:delta_prior} shows the different importance sampling results for different priors. We can see that NH is favored in all the four priors. Values of the logarithm of Beyesian factor for different priors and for the difference between the three new priors case and flat linear priors on $\Delta$ are shown in Table.~\ref{tab:posterpriors}. Adopting the Beyesian model anlysis, we know that flat linear prior on $\Delta$ and  the flat logarithmic prior on $\Sigma m_{\nu}$ are the most preferred prior. Similar to the results in Ref.\cite{Li:2017iur}, priors can change the limits of the sum of neutrino mass, but have little influence on other parameters irrelevant to the neutrino  masses.

\begin{table}[t]
\centering
\begin{tabular}{cccccc}
\hline\hline
 Prior && $\ln{B}$ && $\Delta \ln{B_{ij}}$  \\ 
\hline
flat-linear $\Delta$    &&  -6920.08     && 0.00    \\
flat-log $\Delta$       &&  -6922.42     && -2.34    \\
flat-linear $\Sigma m$  &&  -6922.37     && -2.29    \\
flat-log $\Sigma m$     &&  -6920.08     && 0.00    \\
\hline\hline 
\end{tabular}
\caption{Values of the logarithm of Beyesian factor for different priors and for the difference between the three priors and flat linear priors $\Delta$. }
\label{tab:posterpriors}
\end{table}

\section{Conclusion}
\label{sec:conclu}
In this paper, one available parameter $\Delta$ is adopted to measure the neutrino mass ordering, the positive sign of which represents the normal hierarchy and the negative sign represents the inverted hierarchy. 
We use global observational data, which include cosmology, neutrino oscillation and neutrinoless double beta decay, to constraint neutrino mass and mass hierarchy. 
By comparing about the lightest neutrino mass, the sum of neutrino masses and hierarchy parameter between the results of cosmological data alone and cosmological data +  neutrino oscillation +$0\nu\beta\beta$, we find most of the posterior distribution of hierarchy parameter falls into the $\Delta > 0$ ranges, and a little in the $\Delta < 0$ intervals, which means NH is preferred than IH by the  global data and $\Delta = 0 $ is also strongly disfavored. 
On another hand, $\Sigma m_{\nu} =0.082^{+0.004}_{-0.024}\text{eV} $, which is less than the minimal value of the total neutrino masses in IH case, suggests IH case is ruled out.
For the neutrino mass, the lightest neutrino mass is no more than $0.035 \text{eV}$ and the sum of neutrino mass is no more than $0.133 \text{eV}$ at $95\%$ confidence level. 
Under our method, the combination of cosmological data + neutrino oscillation +$0\nu\beta\beta$ data is strongly favour normal hierarchy.
And our constraints results on the neutrino mixing parameters are tighter  than the results of reference \cite{Hannestad:2007tu}, and consistent with the result of which use a global dataset includes the \textit{Planck2015} CMB data, neutrino oscillation and neutrinoless double beta decay data \cite{Gariazzo:2018pei} by Beyasian analysis.
We also compare the influence of different $\Delta$ priors on all the free parameters we considered by adopting importance sampling technique. 
The three different priors we considered in this paper are the flat logarithmic prior on the absolute value of the neutrino hierarchy parameter $\Delta$, the flat linear prior on the total neutrino mass $\Sigma m_{\nu}$, and the flat logarithmic prior on $\Sigma m_{\nu}$. 
The results show that the priors do not change the preference for the NH. 
Finally, by using the publicly available Beyesian model analysis code \textbf{MCEvidence}, we calculated the Bayes factor of different priors using our MCMC sampling chains, the Bayes factor shows that flat linear prior on $\Delta$ and  the flat logarithmic prior on $\Sigma m_{\nu}$ are the most preferred priors.

\acknowledgments

L.X is supported by National Natural Science Foundation of China under Grant No. 11675032 (People's Republic of China).

\appendix

\section{global data}
\label{app:data}

\subsection{neutrinoless double beta decay data}

The likelihood of neutrinoless double beta decay is following the reference \cite{Caldwell:2017mqu, Gariazzo:2018pei}, the GERDA \cite{Agostini:2017iyd}, KamLAND-Zen \cite{KamLAND-Zen:2016pfg} and EXO-200 \cite{Albert:2014awa} experiments are taken into account:
\begin{align}
\chi^2_{0\nu\beta\beta,\ \rm {GERDA}}(T_{1/2}^{\rm Ge})&=
\frac{(1/T_{1/2}^{\rm Ge}+1.48)^2}{0.461^2}
\,,\label{eq:gerda1}\\
\chi^2_{0\nu\beta\beta,\ \rm {KamLAND-Zen\ phase\ I}}(T_{1/2}^{\rm Xe})&=
2\times \left(2.3/T_{1/2}^{\rm Xe}+1.09/(T_{1/2}^{\rm Xe})^2\right)
\,,\label{eq:lamland11}\\
\chi^2_{0\nu\beta\beta,\ \rm{KamLAND-Zen\ phase\ II}}(T_{1/2}^{\rm Xe})&=
2\times \left(9.71/T_{1/2}^{\rm Xe}+28.1/(T_{1/2}^{\rm Xe})^2\right)
\,,\label{eq:kamland22}\\
\chi^2_{0\nu\beta\beta,\ \rm {EXO}}(T_{1/2}^{\rm Xe})&=
\frac{(1/T_{1/2}^{\rm Xe}-0.32)^2}{0.30^2}
\,,\label{eq:exo1}\\
\chi^2_{0\nu\beta\beta}&=\sum\limits_{i}\chi^2_{0\nu\beta\beta,i} 
\,.\label{eq:0nu}
\end{align}

All the parametrizations have been cross-checked by the experimental collaborations, and the half-life time of neutrinoless double beta decay are given in units of $10^{25}$ years.

\subsection{neutrino oscillation}
Considering the parameters, in our calculations, we can divide the total neutrino oscillation likelihood function $\mathcal{L}_{\rm osc}\propto e^{-\chi^2/2}$ into five parts, then we get the $\chi^2$
 \begin{eqnarray}
\chi^2_{osc}=\chi^2_{\sin^2 \theta_{12}}+\chi^2_{\sin^2 \theta_{13}}+\chi^2_{\sin^2 \theta_{23}}+\chi^2_{log_{10}\Delta m^2_{21}}+\chi^2_{\Delta m^2_{31}}
\end{eqnarray}

We use the latest neutrino oscillation global fit results v3.1, which are taken from \url{nu-fit.org.} \cite{Esteban:2016qun}. Results from solar experiments include the updated external information of the new generation of standard solar models \cite{Vinyoles:2016djt}, the total rate from the radiochemical experiments Chlorine \cite{Cleveland1991Measurement}, Gallex/GNO \cite{Kaether:2010ag} and SAGE \cite{Abdurashitov:2009tn}, the data of the four phases of
Super-KamLAND presented in \cite{Hosaka:2005um, Cravens:2008aa, Abe:2010hy, Nakano:skIV, Nakano:2016ufy}, the results of SNO combinaed analysis \cite{Aharmim:2011vm}, Phase-\uppercase\expandafter{\romannumeral1} and Phase-\uppercase\expandafter{\romannumeral2}of Borexino \cite{Bellini:2011rx, Bellini:2008mr, Bellini2014Neutrinos}.
In the analysis of atmospheric data, the atmospheric neutrino fluxes of external information data \cite{Honda:2015fha} and the results of IceCube/DeepCore \cite{Aartsen:2014yll, icecubeicecube} are included.
Results from the reactor experiments include the updated data of KamLAND \cite{Gando:2013nba} with Daya-Bay \cite{An:2016srz}, Double-Chooz \cite{serra2016double}, Daya-Bay \cite{An:2016ses} and RENO \cite{Seo:2017ksq}.
For the analysis of accelerator experiments, we include the final energy distribution events from MINOS \cite{Adamson:2013whj, Adamson:2013ue}, in $\nu_\mu$ and $\bar{\nu_\mu}$ disappearance and $\nu_\mu$ and $\bar{\nu_\mu}$ appearance channels, as well as from updated T2K \cite{izmaylov2017t2k} in the same four channels and NO$\nu$A \cite{vahle2017new} in $\nu_\mu$ disappearance and $\nu_e$ appearance channels.

\subsection{cosmological data}

The cosmological constraints are obtained from the publicly available Monte-Carlo Markov-Chain package \textbf{CosmoMC} \cite{Lewis:2002ah, Lewis:1999bs}, with a convergence diagnostic based on the Gelman and Rubin statistic, which implements an efficient sampling of the posterior distribution using the fast/slow parameter decorrelations \cite{Lewis:2013hha}, and that includes the support for the Planck data release 2015 Likelihood Code \cite{Aghanim:2015xee}.
The observational data sets we use in this work are composed of CMB, BAO, SN and $H_0$.
 \begin{itemize}
        \item The CMB data: the CMB measurements include the full \textit{Planck} 2015 data release of LowTEB, TT temperature spectrum, EE and TE polarization spectra at whole multiples, provided by the Planck collaboration \cite{Aghanim:2015xee}.
        \item The BAO data: we use the BAO measurements from from 6dFGS \cite{Beutler:2011hx}, SDSS MGS \cite{Ross:2014qpa}, CMASS and LOWZ samples of BOSS DR12 \cite{Gil-Marin:2015nqa}, and also the RSD data from the CMASS and LOWZ \cite{Gil-Marin:2015nqa};
        \item The SN data: for the type Ia supernova observation, we adopt the "Joint Light-curve Analysis"(JLA) sample \cite{Betoule:2014frx};
        \item The $H_0$ data: we employ the result of Riess et al., which is confirmed and improved from their former determination, with the measurement value $H_0 = 73.03 \pm 1.79 {\rm km s}^{-1} {\rm Mpc}^{-1}$  \cite{Riess:2016jrr}.
    \end{itemize}

%%===================================
%\bibliographystyle{JHEP}
%\bibliography{reference.bib}
%\bibliography{referencesss.bib}
%%===================================

\providecommand{\href}[2]{#2}\begingroup\raggedright\endgroup
\end{document}